\documentclass[pre,twocolumn,showpacs,preprintnumbers,amsmath,amssymb]{revtex4}

\usepackage{graphicx}
\usepackage{dcolumn}
\usepackage{bm}

\begin{document}
\title{Phase separation of a driven granular gas in
annular geometry}
\author{Manuel D{\'{\i}}ez-Minguito}
 \affiliation{Institute `Carlos I' for Theoretical and Computational
Physics, University of Granada, E-18071 - Granada, Spain}
\author{Baruch Meerson}
 \affiliation{Racah Institute of Physics, Hebrew University of
Jerusalem, Jerusalem 91904, Israel}
\date{\today}

\begin{abstract}
This work investigates phase separation of a monodisperse gas of inelastically
colliding hard disks confined in a two-dimensional annulus, the inner circle of
which represents a ``thermal wall". When described by granular hydrodynamic
equations, the basic steady state of this system is an azimuthally symmetric
state of increased particle density at the exterior circle of the annulus. When
the inelastic energy loss is sufficiently large, hydrodynamics predicts
spontaneous symmetry breaking of the annular state, analogous to the van der
Waals-like phase separation phenomenon previously found in a driven granular gas
in rectangular geometry. At a fixed aspect ratio of the annulus, the phase
separation involves a ``spinodal interval" of particle area fractions, where the
gas has negative compressibility in the azimuthal direction. The heat conduction
in the azimuthal direction tends to suppress the instability, as corroborated by
a marginal stability analysis of the basic steady state with respect to small
perturbations. To test and complement our theoretical predictions we performed
event-driven molecular dynamics (MD) simulations of this system. We clearly
identify the transition to phase separated states in the MD simulations, despite
large fluctuations present, by measuring the probability distribution of the
amplitude of the fundamental Fourier mode of the azimuthal spectrum of the
particle density. We find that the instability region, predicted from
hydrodynamics, is always located within the phase separation region observed in
the MD simulations. This implies the presence of a binodal (coexistence) region,
where the annular state is metastable. The phase separation persists when the
driving and elastic walls are interchanged, and also when the elastic wall is
replaced by weakly inelastic one.
\end{abstract}


\maketitle

\section{\label{SecI}Introduction}
Flows of granular materials 
are ubiquitous in nature and technology \cite{jaeger}. Examples are numerous and
range from Saturn's rings to powder processing. Being dissipative and therefore
intrinsically far from thermal equilibrium, granular flows exhibit a plethora of
pattern forming instabilities \cite{ristow,aranson+tsimring}. In spite of a
surge of recent interest in granular flows, their quantitative modeling remains
challenging, and the pattern forming instabilities provide sensitive tests to
the models. This work focuses on the simple model of \textit{rapid} granular
flows, also referred to as granular gases: large assemblies of inelastically
colliding hard spheres
\cite{campbell,kadanoff,thorsten1,thorsten2,goldhirsch1,brilliantov}. In the
simplest version of this model the only dissipative effect taken into account is
a reduction in the relative normal velocity of the two colliding particles,
modeled by the coefficient of normal restitution, see below. Under some
additional assumptions a hydrodynamic description of granular gases becomes
possible. The Molecular Chaos assumption allows for a description in terms of
the Boltzmann or Enskog equations, properly generalized to account for the
inelasticity of particle collisions, followed by a systematic derivation of
hydrodynamic equations \cite{sela,brey1,lutsko}. For inhomogeneous (and/or
unsteady) flows hydrodynamics demands scale separation: the mean free path of
the particles (respectively, the mean time between two consecutive collisions)
must be much less than any characteristic length (respectively, time) scale that
the hydrodynamic theory attempts to describe. The implications of these
conditions can be usually seen only \textit{a posteriori}, after the
hydrodynamic problem in question is solved, and the hydrodynamic length/time
scales are determined.  We will restrict ourselves in this work to nearly
elastic collisions and moderate gas densities where, based on previous studies,
hydrodynamics is expected to be an accurate leading order theory
\cite{campbell,kadanoff,thorsten1,thorsten2,goldhirsch1,brilliantov}. These
assumptions allow for a detailed quantitative study (and prediction) of a
variety of pattern formation phenomena in granular gases. One of these phenomena
is the phase separation instability, first predicted in Ref. \cite{livne1} and
further investigated in Refs.
\cite{argentina,brey2,khain1,livne2,baruch2,khain2}. This instability arises
already in a very simple, indeed prototypical setting: a monodisperse granular
gas at zero gravity confined in a rectangular box, one of the walls of which is
a ``thermal" wall. The basic state of this system is the stripe state. In the
hydrodynamic language it represents a laterally uniform stripe of increased
particle density  at the wall opposite to the driving wall. The stripe state was
observed in experiment \cite{kudrolli}, and this and similar settings have
served for testing the validity of quantitative modeling
\cite{kadanoff2,esipov,grossman}. It turns out that (i) within a ``spinodal"
interval of area fractions and (ii) if the system is sufficiently wide in the
lateral direction, the stripe state is unstable with respect to small density
perturbations in the lateral direction \cite{livne1,brey2,khain1}. Within a
broader ``binodal" (or coexistence) interval the stripe state is stable to small
perturbations, but unstable to sufficiently large ones \cite{argentina,khain2}.
In both cases the stripe gives way, usually via a coarsening process, to
coexistence of dense and dilute regions of the granulate (granular ``droplets"
and ``bubbles") along the wall opposite to the driving wall
\cite{argentina,livne2,khain2}. This far-from-equilibrium phase separation
phenomenon is strikingly similar to a gas-liquid transition as described by the
classical van der Waals model, except for large fluctuations observed in a broad
region of aspect ratios around the instability threshold \cite{baruch2}. The
large fluctuations have not yet received a theoretical explanation.

This work addresses a phase separation process in a different geometry. We will
deal here with an assembly of hard disks at zero gravity, colliding
inelastically inside a two-dimensional annulus. The interior wall of the annulus
drives the granulate into a non-equilibrium steady state with a
(hydrodynamically) zero mean flow. Particle collisions with the exterior wall
are assumed elastic. The basic steady state of this system, as predicted by
hydrodynamics, is the \textit{annular} state: an azimuthally symmetric state of
increased particle density at the exterior wall. The phase separation
instability manifests itself here in the appearance of dense clusters with
broken azimuthal symmetry along the exterior wall. Our main objectives are to
characterize the instability and compute the phase diagram by using granular
hydrodynamics (or, more precisely, granular hydrostatics, see below) and event
driven molecular dynamics simulations. By focusing on the annular geometry, we
hope to motivate experimental studies of the granular phase separation which may
be advantageous in this geometry. The annular setting avoids lateral side walls
(with an unnecessary/unaccounted for energy loss of the particles). Furthermore,
driving can be implemented here by a rapid rotation of the (slightly eccentric
and possibly rough) interior circle.

We organized the paper as follows. Section \ref{SecII} deals with a hydrodynamic
description of the annular state of the gas. As we will be dealing only with
states with a zero mean flow, we will call the respective equations hydrostatic.
A marginal stability analysis predicts a spontaneous symmetry breaking of the
annular state. We compute the marginal stability curves and compare them to the
borders of the spinodal (negative compressibility) interval of the system. In
Section \ref{SecIII} we report event-driven molecular dynamics (MD) simulations
of this system and compare the simulation results with the hydrostatic theory.
In Section \ref{SecIV} we discuss some modifications of the model, while Section
\ref{SecV} contains a summary of our results.

\section{\label{SecII} Particles in an annulus and granular hydrostatics}
\textit{The density equation.} Let $N$ hard disks of diameter $d$ and mass $m=1$
move, at zero gravity, inside an annulus of aspect ratio $\Omega=
R_{\text{ext}}/R_{\text{int}}$, where $R_{\text{ext}}$ is the exterior radius
and $R_{\text{int}}$ is the interior one. The disks undergo inelastic collisions
with a constant coefficient of normal restitution
$\mu$. For simplicity, we neglect the rotational degree of freedom of the particles. 
The (driving) interior wall is modeled by a thermal wall kept at temperature
$T_{0}$, whereas particle collisions with the exterior wall are considered
elastic. The energy transferred from the thermal wall to the granulate
dissipates in the particle inelastic collisions, and we assume that the system
reaches a (non-equilibrium) steady state with a zero mean flow. We restrict
ourselves to the nearly elastic limit by assuming a restitution coefficient
close to, but less than, unity: $1-\mu \ll 1$. This allows us to safely use
granular hydrodynamics \cite{goldhirsch1}.  For a zero mean flow steady state
the continuity equation is obeyed trivially, while the momentum and energy
equations yield two \textit{hydrostatic} relations:
\begin{equation}
\nabla \cdot \mathbf{q} (\mathbf{r})+I=0\,, \; \; \; p=const\,,
\label{en_balance_prev}
\end{equation}
where $\mathbf{q}$ is the local heat flux, $I$ is the energy loss term due to
inelastic collisions, and $P=P(n,T)$ is the gas pressure that depends on the
number density $n(\mathbf{r})$ and granular temperature $T(\mathbf{r})$. We
adopt the classical Fourier relation for the heat flux
$\mathbf{q}(\mathbf{r})=-\kappa \nabla T(\mathbf{r})$ (where $\kappa$ is the
thermal conductivity), omitting a density gradient term.  In the dilute limit
this term was derived in Ref. \cite{brey1}. It can be neglected in the nearly
elastic limit which is assumed throughout this paper.

The momentum and energy balance equations read
\begin{equation}
\nabla \cdot \left[ \kappa \nabla T(\mathbf{r}) \right] =I\,, \; \; \;
p=const\,, \label{en_balance}
\end{equation}
To get a closed formulation, we need constitutive relations for $p(n,T)$,
$\kappa(n,T)$ and $I(n,T)$. We will employ the widely used semi-empiric
transport coefficients derived by Jenkins and Richman \cite{jenkins} for
moderate densities:
\begin{equation}
\begin{split}
\kappa=\frac{2d n T^{1/2} \tilde{G}}{\pi^{1/2}}
\left[ 1+\frac{9\pi}{16} \left( 1+\frac{2}{3\tilde{G}}\right)^{2}\right],\\
I=\frac{8(1-\mu)n T^{3/2}\tilde{G}}{d\sqrt{\pi}}\,,
\end{split}
\label{JR}
\end{equation}
and the equation of state first proposed by Carnahan and Starling \cite{carnahan}
\begin{equation}
p=n T(1+2\tilde{G})\,,
\label{CS}
\end{equation}
where $\tilde{G}=\nu (1-\frac{7\nu}{16})/(1-\nu)^{2}$ and $\nu=n \left( \pi
d^{2}/4 \right) $ is the solid fraction. Let us rescale the radial coordinate by
$R_{\text{int}}$ and introduce the rescaled inverse density
$Z(r,\theta)=n_{c}/n(r,\theta)$, where $n_{c}=2/\left( \sqrt{3}d^2 \right)$ is
the hexagonal close packing density. The rescaled radial coordinate $r$ now
changes between $1$ and $\Omega\equiv R_{\text{ext}}/R_{\text{int}}$, the aspect
ratio of the annulus. As in the previous work \cite{khain1}, Eqs.
(\ref{en_balance}), (\ref{CS}) and (\ref{JR}) can be transformed into a single
equation for the inverse density $Z(r)$:
\begin{equation}
\nabla \cdot \left[ \mathcal{F}(Z)\nabla Z\right] =\Lambda \mathcal{Q}(Z)\,,
\label{eq:density_eq}
\end{equation}%
where
\begin{equation}
\begin{split}
\mathcal{F}(Z)=\mathcal{F}_{1}(Z)\mathcal{F}_{2}(Z),\\
\mathcal{Q}(Z)=\frac{6}{\pi} \frac{Z^{1/2}\mathcal{G}}{(1+2\mathcal{G})^{3/2}},\\
\mathcal{F}_{1}(Z)=\frac{\mathcal{G}(Z)\left[ 1+{\frac{9\pi}{16}}
\left( 1+\frac{2}{3\mathcal{G}}\right)^2\right]}{Z^{1/2}(1+2\mathcal{G})^{5/2}},\\
\mathcal{F}_{2}(Z)=1+2\mathcal{G}+ \frac{\pi}{\sqrt{3}}\frac{Z\left(
Z+\frac{\pi}{16\sqrt{3}}\right) }
{\left( Z-\frac{\pi}{2\sqrt{3}}\right)^3},\\
\mathcal{G}(Z)=\frac{\pi}{2\sqrt{3}}\frac{\left(
Z-\frac{7\pi}{32\sqrt{3}}\right)} {\left( Z-\frac{\pi}{2\sqrt{3}}\right) ^2}.
\end{split}
\label{funct}
\end{equation}
The dimensionless parameter $\Lambda \equiv (2\pi/3) (1-\mu)
\left(R_{\text{int}}/d\right)^2$ is the hydrodynamic inelastic loss parameter.
The boundary conditions for Eq.~(\ref{eq:density_eq}) are
\begin{equation}
\partial Z(1,\theta)/\partial\theta=0 \quad \text{and} \quad \nabla_{n} Z(\Omega,\theta)=0\,,
\label{eq:boundary}
\end{equation}
The first of these follows from the constancy of the temperature at the
(thermal) interior wall which, in view of the constancy of the pressure in a
steady state, becomes constancy of the density. The second condition demands a
zero normal component of the heat flux at the elastic wall. Finally, working
with a fixed number of particles, we demand the normalization condition
\begin{equation}
\int_{0}^{2\pi} \int_{1}^{\Omega}  Z^{-1} (r,\theta) r dr d\theta  = \pi f
(\Omega^2-1)\,, \label{eq:norma}
\end{equation}
where $$f=\frac{N}{\pi n_{c}R_{\text{int}}^{2}(\Omega^2-1)}$$ is the area
fraction of the grains in the annulus. Equations
(\ref{eq:density_eq})-(\ref{eq:norma}) determine all possible steady state
density profiles, governed by three dimensionless parameters: $f$, $\Lambda$,
and $\Omega$.\\

\begin{figure}
\includegraphics[width=7.5278cm]{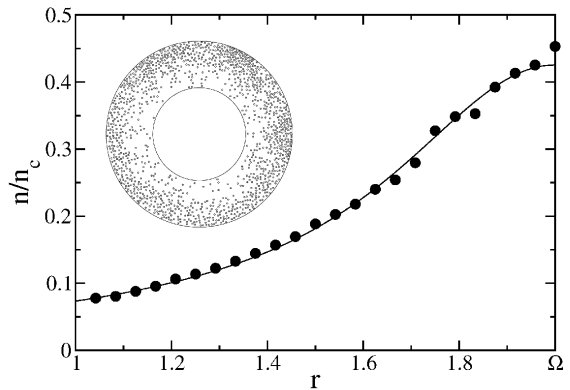}
\caption{\label{fig1} The normalized density profiles obtained from the MD
simulations (the dots) and hydrostatics (the line) for $\Omega=2$,
$\Lambda=81.09$, and $f=0.356$ (equivalently, $z_{\Omega}=2.351$). The
simulations were carried out with $N=1250$ particles, $\mu=0.92$, and
$R_{\text{int}}=22.0$. Also shown is a typical snapshot of the system at the
steady state as observed in the MD simulation.}
\end{figure}

\textit{Annular state.} The simplest solution of the density equation
(\ref{eq:density_eq}) is azimuthally symmetric ($\theta$-independent): $Z=
z(r)$. Henceforth we refer to this basic state of the system as the
\textit{annular state}. It is determined by the following equations:
\begin{equation}
\begin{split}
\left[ r \mathcal{F}(z) z^{\prime} \right]^{\prime} =
r \Lambda \mathcal{Q}(z),\; z^{\prime}(\Omega)=0, \; \text{and} \\
\int_{1}^{\Omega} z^{-1} r dr=(\Omega^{2}-1)f/2 \,,
\end{split}
\label{eq:annular}
\end{equation}
where the primes denote $r$-derivatives. In order to solve the second order
equation (\ref{eq:annular}) numerically, one can prescribe the inverse density
at the elastic wall $z_{\Omega}\equiv z(\Omega)$. Combined with the no-flux condition at
$r=\Omega$, this condition define a Cauchy problem for $z(r)$
\cite{livne2,khain1}. Solving the Cauchy problem, one can compute the respective
value of $f$ from the normalization condition in Eq.~(\ref{eq:annular}). At
fixed $\Lambda$ and $\Omega$, 
there is a one-to-one relation between $z_{\Omega}$ and $f$. Therefore, an alternative parameterization of the
annular state is given by the scaled numbers $z_{\Omega}$, $\Lambda$, and
$\Omega$. The same is true for the marginal stability analysis performed in the
next subsection.

Figure~\ref{fig1} depicts an example of annular state that we found numerically.
One can see that the gas density increases with the radial coordinate, as
expected from the temperature decrease via inelastic losses, combined with the
constancy of the pressure throughout the system. The hydrodynamic density
profile agrees well with the one found in our MD simulations, see below.

\begin{figure}
\includegraphics[width=7.5278cm]{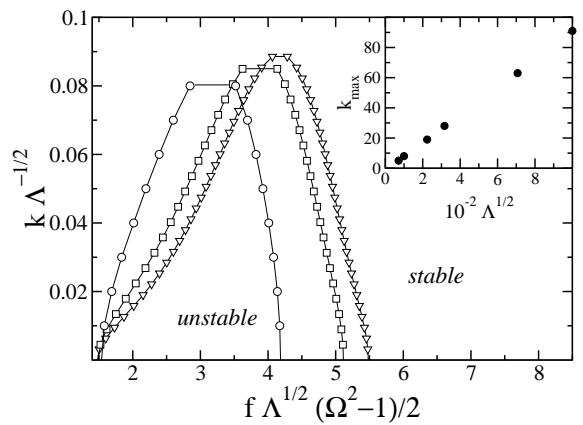}
\caption{\label{fig2} The main graph: the marginal stability curves $k=k(f)$
(where $k$ is an integer) for $\Omega=1.5$ and $\Lambda=10^4$ (circles),
$\Lambda=5\times 10^4$ (squares), and $\Lambda=10^5$ (triangles). For a given
$\Lambda$ the annular state is stable above the respective curve and unstable
below it, as indicated for $\Lambda=10^4$. As $\Lambda$ increases the
marginal stability interval shrinks. 
The inset: the dependence of $k_{max}$ on $\Lambda^{1/2}$.  The straight line
shows that, at large $\Lambda$, $k_{max}\propto \Lambda^{1/2}$.}
\end{figure}

\textit{Phase separation.} Mathematically, phase separation manifests itself in
the existence of \textit{additional} solutions to
Eqs.~(\ref{eq:density_eq})-(\ref{eq:norma}) in some region of the parameter
space $f$, $\Lambda$, and $\Omega$. These additional solutions are \textit{not}
azimuthally symmetric. Solving Eqs.~(\ref{eq:density_eq})-(\ref{eq:norma}) for
fully two-dimensional solutions is not easy \cite{livne1}. One class of such
solutions, however, bifurcate continuously from the annular state, so they can
be found by linearizing Eq.~(\ref{eq:density_eq}), as in rectangular geometry
\cite{livne1,khain1}. In the framework of a time-dependent hydrodynamic
formulation, this analysis corresponds to a \textit{marginal stability} analysis
which involves a small perturbation to the annular state. For a single azimuthal
mode $\sim \sin(k\theta)$ (where $k$ is integer) we can write
$Z(r,\theta)=z(r)+\varepsilon\; \Xi(r) \sin(k\theta)$, where $\Xi(r)$ is a
smooth function, and $\varepsilon \ll 1$ a small parameter. Substituting this
into Eq.~(\ref{eq:density_eq}) and linearizing the resulting equation yields a
$k$-dependent second order differential equation for the function
$\Gamma(r)\equiv\mathcal{F}[Z(r)]\,\Xi(r)$:
\begin{equation}
\Gamma^{\prime\prime}_k+\frac{1}{r}\Gamma^{\prime}_k- \left(
\frac{k^{2}}{r^{2}}+\frac{\Lambda \mathcal{Q}^{\prime}(Z)}
{\mathcal{F}(Z)}\right) \Gamma_{k}=0\,. \label{eq:MSA}
\end{equation}%
This equation is complemented by the boundary conditions
\begin{equation}
\Gamma(1)=0  \;\;\; \mbox{and} \;\;\; \Gamma^{\prime}(\Omega)=0\,.
\label{eq:MSA_BC}
\end{equation}%
For fixed values of the scaled parameters $f$, $\Lambda$, and $\Omega$,
Eqs.~(\ref{eq:MSA}) and (\ref{eq:MSA_BC}) determine a linear eigenvalue problem
for $k$. Solving this eigenvalue problem numerically, one obtains the marginal
stability hypersurface  $k=k(f,\Lambda,\Omega)$. For fixed $\Lambda$ and
$\Omega$, we obtain a marginal stability curve $k=k(f)$. Examples of such
curves, for a fixed $\Omega$ and three different $\Lambda$ are shown in
Fig.~\ref{fig2}. Each $k=k(f)$ curve has a maximum $k_{max}$, so that a density
modulation with the azimuthal wavenumber larger than $k_{max}$ is stable. As
expected, the instability interval is the largest for the fundamental mode
$k=1$. The inset in Fig.~\ref{fig2} shows the dependence of $k_{max}$ on
$\Lambda^{1/2}$.  The straight line shows that, at large $\Lambda$,
$k_{max}\propto \Lambda^{1/2}$, as in rectangular geometry \cite{khain1}.

\begin{figure}
\includegraphics[width=7.5278cm]{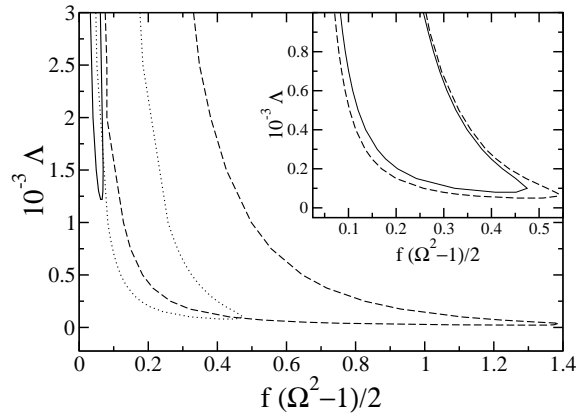}
\caption{\label{fig3} Two-dimensional projections on the $[\Lambda$,
$f\,(\Omega^{2}-1)/2]$ plane of the phase diagram at $\Omega=1.5$ (the solid
line), $\Omega=3$ (the dotted line), and $\Omega=5$ (the dashed line). The inset
shows more clearly, for $\Omega=3$, that the marginal stability curve (the solid
line) lies within the negative compressibility region (bounded by the dashed
line).}
\end{figure}

Two-dimensional projections of the ($f$, $\Lambda$, $\Omega$)-phase diagram at
three different $\Omega$ are shown in Fig.~\ref{fig3} for the fundamental mode.
The annular state is unstable in the region bounded by the marginal stability
curve, and stable elsewhere. Therefore, the marginal stability analysis predicts
loss of stability of the annular state within a finite interval of $f$, that is
at $f_{min}(\Lambda,\Omega)<f<f_{max}(\Lambda,\Omega)$.

\begin{figure}
\includegraphics[width=7.5278cm]{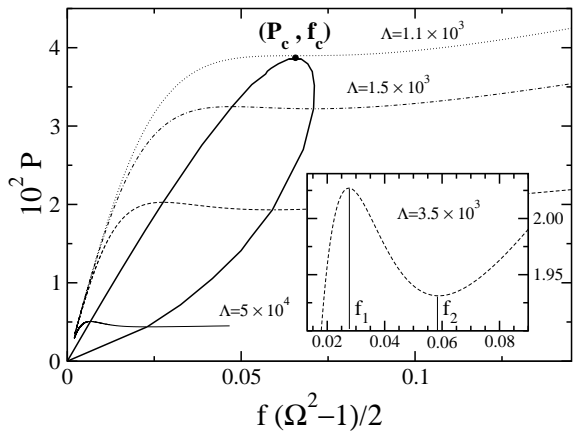}
\caption{\label{fig4} The scaled steady state granular pressure $P$ versus the
grain area fraction $f$ for $\Omega=1.5$ and $\Lambda=1.1\times 10^{3}$ (the
dotted line), $\Lambda=1.5\times 10^{3}$ (the dash-dotted line),
$\Lambda=3.5\times 10^{3}$ (dashed line), and $\Lambda=5\times 10^{4}$ (the
solid line). The inset shows a zoom-in for $\Lambda=3.5\times 10^{3}$. The
borders $f_1$ and $f_2$ of the spinodal interval are determined from the
condition $\left(
\partial P/\partial f \right)_{\Lambda,\Omega}=0$. The thick solid line
encloses the spinodal balloon where the effective azimuthal compressibility of
the gas is negative.}
\end{figure}

The physical mechanism of this phase separation \textit{instability} is the
negative compressibility of the granular gas in the azimuthal direction, caused
by the inelastic energy loss. To clarify this point, let us compute the
\textit{pressure} of the annular state, given by Eq. (\ref{CS}). First we
introduce a rescaled pressure $P=p/(n_{c}T_{0})$ and, in view of the pressure
constancy in the annular state, compute it at the thermal wall, where $T=T_{0}$
is prescribed and $z(1)$ is known from our numerical solution for the annular
state. We obtain
$$
P(f,\Lambda,\Omega)=\frac{1+2\mathcal{G}(z(1))}{z(1)}\,.
$$
The spinodal (negative compressibility) region is determined by the necessary
condition for the \textit{instability}: $\left(
\partial P/\partial f\right)_{\Lambda,\Omega}<0$, whereas the borders of the
spinodal region are defined by $\left( \partial P/\partial
f\right)_{\Lambda,\Omega}=0$. Typical $P(f)$ curves for a fixed $\Omega$ and
several different $\Lambda$ are shown in Fig.~\ref{fig4}. One can see that, at
sufficiently large $\Lambda$, the rescaled pressure $P$ goes down with an
increase of $f$ at an interval $f_{1}<f<f_{2}$. That is, the effective
compressibility of the gas with respect to a redistribution of the material in
the azimuthal direction is negative on this interval of area fractions. By
joining the spinodal points $f_1$ and (separately) $f_2$ at different $\Lambda$,
we can draw the spinodal line for a fixed $\Omega$. As $\Lambda$ goes down, the
spinodal interval shrinks and eventually becomes a point at a critical point
$(P_{c},f_c)$, or $(\Lambda_{c},f_c)$ (where all the critical values are
$\Omega$-dependent). For $\Lambda<\Lambda_c$ $P(f)$ monotonically increases and
there is no instability.

What is the relation between the spinodal interval $(f_{1},f_{2})$ and the
marginal stability interval $(f_{min},f_{max})$? These intervals would coincide
were the azimuthal  wavelength of the perturbation infinite (or, equivalently,
$k\rightarrow 0$), so that the azimuthal heat conduction would vanish. Of
course, this is not possible in annular geometry, where $k\ge 1$. As a result,
the negative compressibility interval must include in itself the marginal
stability interval $(f_{min},f_{max})$. This is what our calculations indeed
show, see the inset of Fig.~\ref{fig3}. That is, a negative compressibility is
necessary, but not sufficient, for instability, similarly to what was found in
rectangular geometry \cite{khain1}.

Importantly, the instability region of the parameter space is by no means not
the \textit{whole} region the region where phase separation is expected to
occur. Indeed, in analogy to what happens in rectangular geometry
\cite{argentina,khain2}, phase separation is also expected in a \textit{binodal}
(or coexistence) region of the area fractions, where the annular state is stable
to small perturbations, but unstable to sufficiently large ones. The whole
region of phase separation should be larger than the instability region, and it
should of course \textit{include} the instability region. Though we did not
attempt to determine the binodal region of the system from the hydrostatic
equations (this task has not been accomplished yet even for rectangular
geometry, except in the close vicinity of the critical point \cite{khain2}), we
determined the binodal region from our MD simulations reported in the next
section.

\section{\label{SecIII}MD Simulations}
\textit{Method}. We performed a series of event-driven MD simulations of this
system using an algorithm described by P\"{o}schel and Schwager
\cite{thorsten3}. Simulations involved $N$ hard disks of diameter $d=1$ and mass
$m=1$. After each collision of particle $i$ with particle $j$, their relative
velocity is updated according to
\begin{equation}
\vec{v}_{ij}^{\,\prime}=\vec{v}_{ij} - \left( 1+\mu \right) \left(
\vec{v}_{ij}\cdot \hat{r}_{ij}\right)\hat{r}_{ij}\,, \label{eq:velocs}
\end{equation}
where $\vec{v}_{ij}$ is the precollisional relative velocity, and
$\hat{r}_{ij}\equiv \vec{r}_{ij}/\left|\vec{r}_{ij}\right|$ is a unit vector
connecting the centers of the two particles. Particle collisions with the
exterior wall $r=R_{\text{ext}}$ are assumed elastic. The interior wall is kept
at constant temperature $T_{0}$ that we set to unity. This is implemented as
follows. When a particle collides with the wall it forgets its velocity and
picks up a new one from a proper Maxwellian distribution with temperature
$T_{0}$, see \textit{e.g}. Ref. \cite{thorsten3}, pages 173-177, for detail.
The time scale is therefore $d(m/T_{0})^{1/2}=1$. The initial condition is a
uniform distribution of non-overlapping particles inside the annular box. Their
initial velocities are taken randomly from a Maxwellian distribution at
temperature $T_{0}=1$. In all simulations the coefficient of normal restitution
$\mu=0.92$ and the interior radius $R_{\text{int}}/d=22.0$ were fixed, whereas
the the number of particles $527\leq N\leq 7800$ and the aspect ratio
$1.5\leq\Omega\leq 6$ were varied. In terms of the three scaled hydrodynamic
parameters the heat loss parameter $\Lambda=81.09$ was fixed whereas $f$ and
$\Omega$ varied.

To compare the simulation results with predictions of our hydrostatic theory,
all the measurements were performed once the system reached a steady state. This
was monitored by the evolution of the total kinetic energy $(1/2)
\sum_{i=1}^{N}\vec{v}_{i}^{\;2}$, which first decays and then, on the
average,  stays constant.\\

\begin{figure}
\includegraphics[width=7.5278cm]{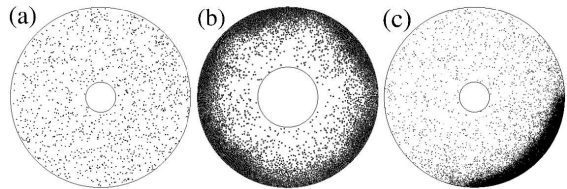}
\caption{\label{fig5} Typical steady state snapshots for $N=1250$ and $\Omega=6$
(a); $N=5267$ and $\Omega=3$ (b), and $N=6320$ and $\Omega=6$ (c). Panels (a)
and (b) correspond to annular states of the hydrostatic theory, whereas panel
(c) shows a broken-symmetry (phase separated) state.}
\end{figure}

\textit{Steady States}. Typical steady state snapshots of the system, observed
in our MD simulation, are displayed in Fig.~\ref{fig5}. Panel (a) shows a dilute
state where the radial density inhomogeneity, though actually present, is not
visible by naked eye. Panels (b) and (c) do exhibit a pronounced radial density
inhomogeneity. Apart from visible density fluctuations, panels (a) and (b)
correspond to annular states. Panel (c) depicts a broken-symmetry (phase
separated) state. When an annular state is observed, its density profile agrees
well with the solution of the hydrostatic equations
(\ref{eq:density_eq})-(\ref{eq:norma}). A typical example of such a comparison
is shown in Fig.~\ref{fig1}.

\begin{figure}
\includegraphics[width=7.5278cm]{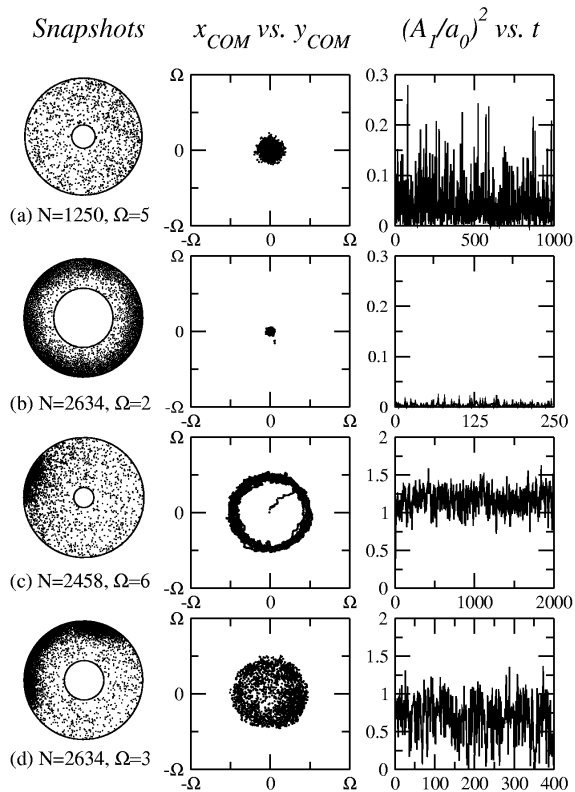}
\caption{\label{fig6} Typical steady state snapshots (the left column) and the
temporal evolution of the COM (the middle column) and of the squared amplitude
of the fundamental Fourier mode (the right column). The temporal data are
sampled each $150$ collisions per particle. Each row corresponds to one
simulation with the indicated parameters. The vertical scale of panels a and b
was stretched for clarity.}
\end{figure}

Let us fix the aspect ratio $\Omega$ of the annulus at not too a small value and
vary the number of particles $N$. First, what happens on a qualitative level?
The simulations show that, at small $N$, dilute annular states, similar to
snapshot (a) in Fig.~\ref{fig5}, are observed. As $N$ increases,
broken-symmetric states start to appear. Well within the unstable region, found
from hydrodynamics, a high density cluster appears, like the one shown in
Fig.~\ref{fig5}c, and performs an erratic motion along the exterior wall. As $N$
is increased still further, well beyond the high-$f$ branch of the unstable
region, an \textit{annular state} reappears, as in Fig.~\ref{fig5}b. This time,
however, the annular state is denser, while its local structure varies from a
solid-like (with imperfections such as voids and line defects) to a liquid-like.

To characterize the spatio-temporal behavior of the granulate at a steady state,
we followed the position of the center of mass (COM) of the granulate. Several
examples of the COM trajectories are displayed in Fig.~\ref{fig6}. Here cases
(a) and (b) correspond, in the hydrodynamic language, to annular  states. There
are, however, significant fluctuations of the COM around the center of the
annulus. These fluctuations are of course not accounted for by hydrodynamic
theory. In case (b), where the dense cluster develops, the fluctuations are much
weaker that in case (a). More interesting are cases (c) and (d). They correspond
to broken-symmetry states: well within the phase separation region of the
parameter space (case c) and close to the phase separation border (case d). The
COM trajectory in case (c) shows that the granular ``droplet" performs random
motion in the azimuthal direction, staying close to the exterior wall. This is
in contrast with case (d), where fluctuations are strong both in the azimuthal
and in the radial directions. Following the actual snapshots of the simulation,
one observes here a very complicated motion of the ``droplet", as well as its
dissolution into more ``droplets", mergers of the droplets \textit{etc}.
Therefore, as in the case of granular phase separation in rectangular geometry
\cite{baruch2}, the granular phase separation in annular geometry is accompanied
by considerable spatio-temporal fluctuations. In this situation a clear
distinction between a phase-separated state and an annular state, and a
comparison between the MD simulations and hydrodynamic theory, demand proper
diagnostics. We found that such diagnostics are provided by the azimuthal
spectrum of the particle density and its probability distribution.

\begin{figure}
\includegraphics[width=7.5278cm]{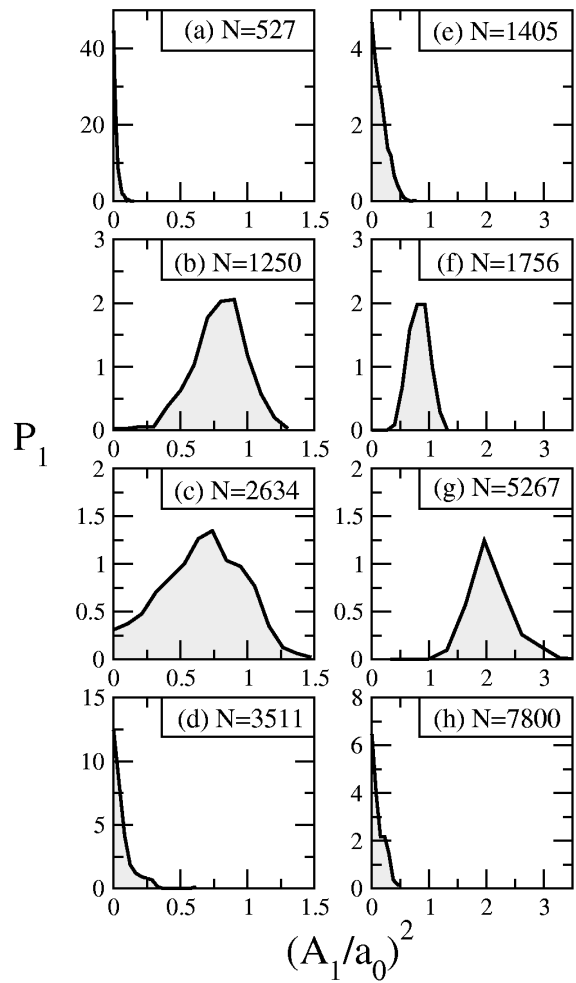}
\caption{\label{fig7} The normalized probability distribution functions
$P_{1}\left(A_{1}^{2}/a_{0}^{2} \right)$ for $\Omega=3$ (the left column) and
$\Omega=5$ (the right column) for different numbers of particles.}
\end{figure}

\textit{Azimuthal Density Spectrum}. Let us consider the (time-dependent)
rescaled density field $\nu(r,\theta,t)=n(r,\theta,t)/n_c $ (where $r$ is
rescaled to the interior wall radius as before), and introduce the integrated
field $\hat{\nu}(\theta, t)$:
\begin{equation}
\hat{\nu}(\theta,t) = \int_{1}^{\Omega}\,\nu(r,\theta,t)\, r \, dr \,.
\end{equation}
In a system of $N$ particles, $\hat{\nu}(\theta,t)$ is normalized so that
\begin{equation}\label{nunorm}
    \int_{0}^{2 \pi} \hat{\nu} (\theta,t)\, d\theta=\frac{N}{n_c\,R^2_{\text{int}}}\,.
\end{equation}
Because of the periodicity in $\theta$ the function $\hat{\nu}(\theta,t)$ can be
expanded in a Fourier series:
\begin{equation}
\hat{\nu}(\theta,t) = a_0+\sum_{k=1}^{\infty} \left[ a_k (t) \cos (k\theta) +
b_k(t) \sin (k\theta)\right] \,,
\end{equation}
where $a_0$ is independent of time because of the normalization condition
(\ref{nunorm}). We will work with the quantities
\begin{equation}
A_k^{2}(t)\equiv a_k^{2}(t) + b_k^{2}(t) \,, \;\;\;k\ge1\,. \label{eq:spectral}
\end{equation}
For the (deterministic) annular state one has $A_k=0$ for all $k\ge 1$, while
for a symmetry-broken state $A_k>0$. The relative quantities $A_k^2(t)/a_{0}^2$
can serve as measures of the azimuthal symmetry breaking. As is shown in
Table~\ref{tab:Ak}, $A_1^2(t)$ is usually much larger (on the average) that the
rest of $A_{k}^{2}(t)$. Therefore, the quantity $A_1^2(t)/a_{0}^2$ is sufficient
for our purposes.

\begin{table}
\caption{\label{tab:Ak}The averaged squared relative amplitudes $\langle
A_{k}^2(t)\rangle/a_{0}^2$ for the first three modes $k=1,2$ and $3$. (a)
$N=2634$, $\Omega=3$, (b) $N=5267$, $\Omega=4$, (c) $N=1000$, $\Omega=2.25$, and
(d) $N=1250$, $\Omega=3$.}
\begin{ruledtabular}
\begin{tabular}{ccccc}
$k$&(a)&(b)&(c)&(d)\\
\hline
$1$ & $0.66\pm 0.05$ & $0.39\pm 0.04$ & $0.30\pm 0.08$ & $0.77\pm 0.05$\\
$2$ & $0.04\pm 0.02$ & $0.05\pm 0.02$ & $0.07\pm 0.01$ & $0.28\pm 0.09$\\
$3$ & $0.03\pm 0.02$ & $0.03\pm 0.03$ & $0.02\pm 0.02$ & $0.11\pm 0.08$\\
\end{tabular}
\end{ruledtabular}
\end{table}

Once the system relaxed to a steady state, we followed the temporal evolution of
the quantity $A_{1}^{2}/a_0^2$. Typical results are shown in the right column of
Fig.~\ref{fig6}. One observes that, for annular states, this quantity is usually
small, as is the cases (a) and (b) in Fig.~\ref{fig6}. For broken-symmetry states
$A_{1}^{2}$ is larger, and it increases as one moves deeper into the phase
separation region. (Notice that in Fig.~\ref{fig6} the averaged value
of $A_{1}^{2}/a_0^2$ in (c) is
larger than in (d), which means that (c) is deeper in the phase separation
region.) Another characteristics of $A_{1}^{2}(t)/a_0^2$ is the magnitude of
fluctuations. One can notice that, in the vicinity of the phase separation
border the fluctuations are stronger (as in case (d) in Fig.~\ref{fig6}).

All these properties are encoded in the \textit{probability distribution}
$P_{1}$ of the values of $\left(A_{1}/a_{0}\right)^{2}$: the ultimate tool of
our diagnostics. Figure~\ref{fig7} shows two series of measurements of this
quantity at different $N$: for $\Omega=3$ and $\Omega=5$. By following the
position of the maximum of $P_{1}$ we were able to to sharply discriminate
between the annular states and phase separated states and therefore to locate
the phase separation border. When the maximum of $P_1$ occurs at the zero value
of $\left(A_{1}/a_{0}\right)^{2}$ (as in cases (a) and (d) and, respectively, (e) and
(h) in Fig.~\ref{fig7}), an annular state is observed. On the contrary, when the
maximum of $P_1$ occurs at a non-zero value of $\left(A_{1}/a_{0}\right)^{2}$
(as in cases (b) and (c) and, respectively, (f) and (g) in Fig.~\ref{fig7}), a phase
separated state is observed. In each case, the width of the probability
distribution (measured, for example, at the half-maximum) yields a direct
measure of the magnitude of fluctuations. Near the phase separation border,
strong fluctuations (that is, broader distributions) are observed, as in case (c)
of Fig.~\ref{fig7}.

\begin{figure}
\includegraphics[width=7.5278cm]{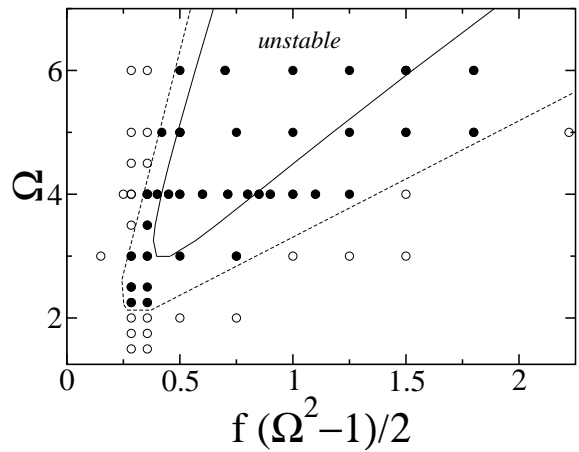}
\caption{\label{fig8} The $\Omega$-$f$ phase diagram for $\Lambda=81.09$. The
solid curve is given by the granular hydrostatics: it shows the borders of the
region where the annular state is unstable with respect to small perturbations.
The filled symbols depict the parameters in which phase separated states are
observed, whereas the hollow symbols show the parameters at which annular states
are observed. The dashed line is an estimated binodal line of the system.}
\end{figure}

Using the position of the maximum of $P_1$ as a criterion for phase separation,
we show, in Fig.~\ref{fig8}, the $\Omega-f$ diagram obtained from the MD
simulations. The same figure also depicts the hydrostatic prediction of the
\textit{instability} region. One can see that the instability region is located
within the phase separation region, as expected.

\begin{figure}
\includegraphics[width=7.5278cm]{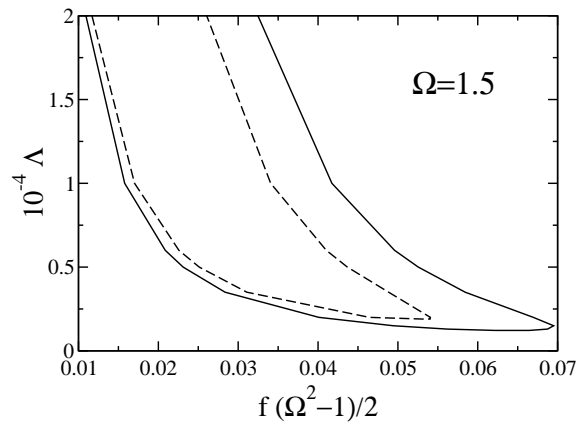}
\caption{\label{fig9} The marginal stability lines for our main setting (the
solid line) and for an alternative setting in which the thermal wall is at
$r=R_{\textit{ext}}$ and the elastic wall is at $r=R_{\textit{int}}$ (the dashed
line).}
\end{figure}

\section{\label{SecIV}Some modifications of the model}
We also investigated an alternative setting in which the exterior wall is the
driving wall, while the interior wall is elastic. The corresponding hydrostatic
problem is determined by the same three scaled parameters $f$, $\Lambda$ and
$\Omega$, but the boundary conditions must be changed accordingly. Here
azimuthally symmetric clusters appear near the (elastic) interior wall. Symmetry
breaking instability occurs here as well. We found very similar marginal
stability curves here, but they are narrower (as shown in Fig.~\ref{fig9}) than
those obtained for the original setting.

\begin{figure}
\includegraphics[width=7.5278cm]{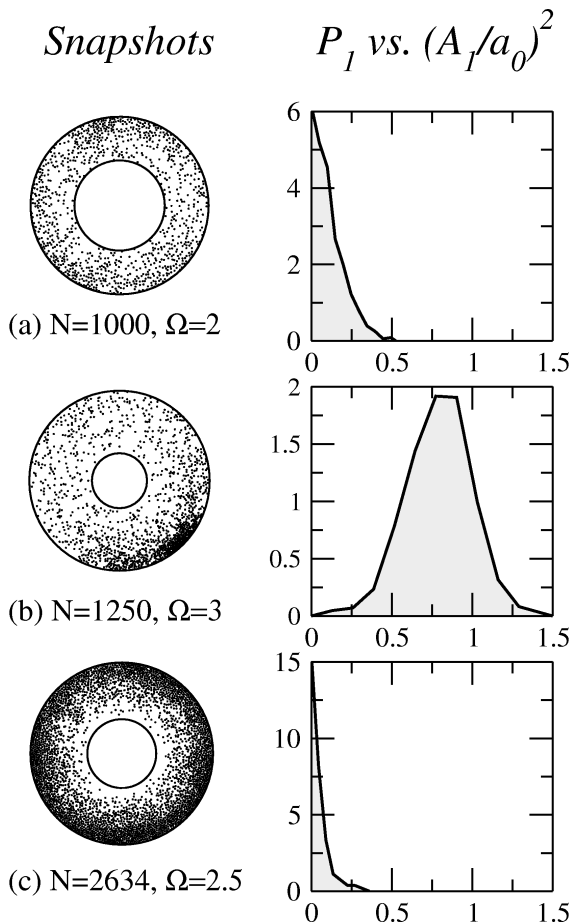}
\caption{\label{fig10} Typical steady state snapshots (the left column) and the
normalized probability distribution functions $P_{1}\left(A_{1}^{2}/a_{0}^{2}
\right)$ for an inelastic exterior wall, $\mu_{wall}=0.99$ (the right column)
for different numbers of particles.}
\end{figure}

Finally, we returned to our original setting and performed several MD
simulations, replacing the perfectly elastic exterior wall by a weakly inelastic
one. The inelastic particle collisions with the exterior wall were modeled in
the same way as the inelastic collisions between particles. Typical results of
these simulations are shown in Fig.~\ref{fig10}. It can be seen that, for the
right choice of parameters, the phase separation persists. This result is
important for a possible experimental test of our theory.

\section{\label{SecV}Summary}
We combined equations of granular hydrostatics and event-driven MD simulations
to investigate spontaneous phase separation of a monodisperse gas of
inelastically colliding hard disks in a two-dimensional annulus, the inner
circle of which serves as a ``thermal wall". A marginal stability analysis
yields a region of the parameter space where the annular state -- the basic,
azimuthally symmetric steady state of the system -- is unstable with respect to
small perturbations which break the azimuthal symmetry. The physics behind the
instability is negative effective compressibility of the gas in the azimuthal
direction, which results from the inelastic energy loss. MD simulations of this
system show phase separation, but it is masked by large spatio-temporal
fluctuations. By measuring the probability distribution of the amplitude of the
fundamental Fourier mode of the azimuthal spectrum of the particle density we
have been able to clearly identify the transition to phase separated states in
the MD simulations. We have found that the instability region of the parameter
space, predicted from hydrostatics, is located within the phase separation
region observed in the MD simulations. This implies the presence of a binodal
(coexistence) region, where the annular state is \textit{metastable}, similar to
what was found in rectangular geometry \cite{argentina,khain2}. The instability
persists in an alternative setting (a driving exterior wall and an elastic
interior wall), and also when the elastic wall is replaced by a weakly inelastic
one. We hope our results will stimulate experimental work on the phase
separation instability.

\section{Acknowledgments}
This work grew out of a student research project at the 2003 Summer School
``From pattern formation to granular physics and soft condensed matter"
sponsored by the European Union under FP5 High Level Scientific Conferences, and
by the NATO Advanced Study Institute. We are grateful to Igor Aranson, Pavel
Sasorov and Thomas Schwager for advice. MDM acknowledges financial support from
MEyC and FEDER (project FIS2005-00791). BM acknowledges financial support from
the Israel Science Foundation (grant No. 107/05) and from the German-Israel
Foundation for Scientific Research and Development (Grant I-795-166.10/2003).

\end{document}